\documentclass[authoryear,1p]{elsarticle}

\begin{document}

\title{Formation of Jupiter using opacities based on detailed grain physics}

\author[ucscpmc,tau]{Naor Movshovitz}
\ead{nmovshov@ucsc.edu}
\author[ucolick]{Peter Bodenheimer}
\author[tau]{Morris Podolak}
\author[ames]{Jack J.~Lissauer}

\address[ucscpmc]{Department of Earth and Planetary Sciences, University 
of California, Santa Cruz, CA 95064, USA}
\address[tau]{Department of Geophysics and Planetary Sciences, Tel Aviv 
University, Ramat Aviv, 69978 Israel}
\address[ucolick]{UCO/Lick Observatory, Department of Astronomy and Astrophysics, University of California, Santa Cruz, CA 95064, USA}
\address[ames]{Space Science and Astrobiology Division,
MS 245-3, NASA-Ames Research Center, Moffett Field, CA 94035, USA}

\begin{abstract}
Numerical simulations, based on the core-nucleated accretion model, are
presented for the formation of Jupiter at 5.2 AU in 3 primordial
disks with three different assumed values of the surface density of 
solid particles. The grain opacities in the envelope of the protoplanet
are computed using a detailed model that includes settling and coagulation
of grains and that incorporates a recalculation of the grain size distribution
at each point in time and space. We generally find lower opacities than the 2\% of interstellar
 values used in previous calculations [Hubickyj, O.,
Bodenheimer, P., Lissauer, J.~J., 2005. Icarus 179, 415--431;
Lissauer, J.~J., Hubickyj, O., D'Angelo, G., Bodenheimer, P., 2009.
Icarus 199, 338-350].  These lower opacities result in more rapid
heat loss from and more rapid contraction of the protoplanetary
envelope. For a given surface density of solids, the new calculations
result in a substantial speedup  in formation time as compared with
those previous calculations.  Formation times are calculated to
be $1.0$, $1.9$, and $4.0$ Myr, and solid core masses are found to be
$16.8$, $8.9$, and $4.7\;\mathrm{M}_{\oplus}$, for solid surface densities, $\sigma$,
of $10$, $6$, and $4\;\mathrm{g\;cm^{-2}}$, respectively. For $\sigma=10$ and
$\sigma=6\;\mathrm{g\;cm^{-2}}$, respectively, these formation times are
reduced by more than 50\% and more than 80\% compared with those
in a previously published calculation with the old approximation
to the opacity.
\end{abstract}


\maketitle

\section{Introduction\label{sect:intro}}
The core-nucleated  accretion theory of giant planet formation was effectively originated by \citet{safronov72}, who realized that a major part of the process involved the accumulation of solid bodies. Consider the accretion of relatively small objects (planetesimals) onto a significantly larger embryo. If $M_{\rm core}$ is the mass of the embryo, then the fundamental equation for its growth, in the absence of gas, is
\begin{equation}
\frac{{\rm d}M_{\rm core}}{{\rm d}t} =  \pi R_{\rm core}^2 \sigma\Omega F_g
\approx \pi R_{\rm core}^2 \sigma\Omega
\bigl[1 + \bigl(\frac{v_e}{v}\bigr)^2\bigr], 
\label{eq:saf}
\end{equation}
where $\pi R_{\rm core}^2$  is the effective geometrical capture 
cross-section, $\sigma$ is the surface density of solid material (planetesimals), 
 $v_e$ is the escape velocity from the embryo, $\Omega$ 
is the orbital frequency, and $v$ is the relative velocity of embryo
and planetesimal. The quantity in brackets is  the
gravitational enhancement factor $F_g$, here given  in the (2 + 2)-body approximation.
 \citeauthor{safronov72} mentioned the later
capture of gas to form giant planets, and \citet{cameron73} stated that
Jupiter could form by gas accretion once a solid core of about 10 M$_\oplus$
had been accreted.  \citet{perri74} then constructed an
adiabatic, hydrostatic equilibrium model of a protoplanet with a core
and a gaseous envelope, extending outward to the Hill radius.  They were
able to show that there was a
critical mass for the core, above which the envelope was unstable,
implying that rapid gas accretion could occur. However, the value of this 
 critical core mass was at least 70 M$_\oplus$,
 much higher than  the deduced core
masses of Jupiter and Saturn. The assumption of adiabaticity turns out
to be incorrect \citep{stevenson82}, as more detailed calculations show
that radiative transport with subadiabatic gradients is able to carry
the energy outwards in significant portions of a growing giant planet.

 \citet{mizuno80} constructed models of a protoplanet 
including  both
radiative and convective energy transport in the envelope, along with 
an energy source provided by the planetesimals accreting onto the core 
at a constant rate, typically  $\dot M_{\rm core} =10^{-6}$ M$_\oplus$/yr.
The model assumed strict hydrostatic and thermal equilibrium, and the
outer radius was assumed to be the protoplanet's Hill radius. Again a critical
core mass was found, above which the envelope was unstable, and the
value was about 12 M$_\oplus$, close to the estimated core masses of
the giant planets. Furthermore, the critical value was practically
independent of the distance to the Sun, partly as a result of the 
assumption of the same constant value of $\dot M_{\rm core}$ at all distances.

The assumption of strict hydrostatic equilibrium was relaxed by 
\citet{bodenheimer86}, who allowed quasi-static contraction
and calculated the first evolutionary sequences applied to the
core-nucleated accretion model. The planetesimal accretion rate
was assumed to be constant, also at $10^{-6}$ M$_\oplus$/yr. They
found that gravitational contraction  (but not dynamical collapse)
began to become important at 
about the point where core and envelope masses were equal, at a value 
which later came to be known as the {\it crossover} mass. Rapid gas
accretion also began near the crossover mass, which they found to be about 
16.5 M$_\oplus$.

The next major step was a series of calculations by \citet{pollack96}
who allowed a variable core accretion rate given by eq.~(\ref{eq:saf}).
The simulations contained three major elements: (1) three-body accretion
cross-sections of solids onto the embryo \citep{greenzweig92} to
determine $F_g$, (2) a modified stellar evolution code \citep{henyey64} 
to follow the structure and evolution of the gaseous envelope, and 
(3) a calculation of the trajectories of planetesimals through the
gaseous envelope \citep{podolak88}, to deduce the deposition of
solid mass and energy in the envelope and to determine the protoplanet's
effective capture radius, taking into account gas drag. The often-quoted
result of these calculations is that it took Jupiter 8 Myr to form in
a solar nebula with $\sigma=10$ g cm$^{-2}$, three times the density in
the minimum mass solar nebula, under the assumption that the opacity
due to small grains in the cool outer envelope was provided by grains
with an interstellar size distribution and with interstellar abundances
with respect to the gas. It was also shown that if the
grain opacities were arbitrarily reduced to 2\% of interstellar values,
the formation time was reduced to about 3 Myr.

Subsequent observations of protoplanetary disks and modelling of the interior of Jupiter revealed two major problems with the 
\citet{pollack96} baseline-case results for the
formation of Jupiter.  First, that   case gave a
formation time longer than the mean lifetime of protostellar disks, 
and, second, it produced a core mass of about  16  M$_\oplus$, 
above the plausible range at the time of  (0--11) M$_\oplus$, 
 derived from matching
interior models of the present Jupiter with observations \citep{saumon04}. Further
calculations \citep{hubickyj05} considered, with generally improved
physics, the effects of the following parameters: (1) the opacity arising
from small grains in the envelope, (2) the solid surface density $\sigma$
in the protoplanetary disk, and (3) a possible cutoff in core accretion
rate as a result of competition for solid particles by neighboring
embryos. The reduction in opacity was justified by calculations \citep{podolak03} showing that grains entering into the protoplanetary envelope
would rapidly coagulate, settle, and eventually evaporate, resulting in an
actual opacity well below interstellar values. In the actual protoplanetary
formation calculation the grain opacity was simply reduced to 2\% of 
interstellar values, with the result that the formation time was reduced
to under     3 Myr, half that found for
interstellar values of grain opacity. 
The final core mass, however, still remained near 
16 M$_\oplus$. A reduction in $\sigma$ from 10 to 6 g cm$^{-2}$, again with
the opacity at 2\% of interstellar, 
increased the formation time to 13 Myr and reduced the core mass to
8 M$_\oplus$. It turned out that the constraints of short formation time and
small core mass could be satisfied simultaneously only if the accretion
rate of solids into the protoplanet were arbitrarily cut off at some time, 
simulating roughly the effect of planetesimal accretion by neighboring
protoplanets. 
A cutoff at $M_{\rm core} = 5$ M$_\oplus$ led to a formation time of
4.5 Myr, while a cutoff at 10 M$_\oplus$ gave a formation time of 1 Myr.

In the planet formation simulations of \citet{pollack96} 
 and \citet{hubickyj05},  simplified
surface boundary conditions were applied during the rapid gas accretion
phase. First, the gas density in the protoplanetary disk was assumed to
be constant in time. Second, the gas accretion rate was capped at
$10^{-2}$ M$_\oplus$/yr, a rough estimate of the rate at which the disk
could supply gas to the protoplanet as a result of its viscous evolution.
Third, the outer boundary of the protoplanet was set at the modified
accretion radius $R_A$, defined by \citet{bodenheimer00} to be 
\begin{equation}
R_A = \frac{GM_p}{c_s^2/k_1 + \frac{GM_p}{(k_2R_H)}},
\label{eq:ra}
\end{equation}
where $R_H$ is the Hill sphere radius, $c_s$ is  the sound speed
in the disk, $M_p$ is the planet's mass, and $k_1,~k_2$ are constants,
both set to unity. Equation~(\ref{eq:ra})  is an interpolation between the Bondi
accretion radius and the Hill sphere radius, such that when $R_H$ is
large, $R_A$ reduces to the Bondi radius, and when $R_H$ is small,
$R_A$ reduces to the Hill radius. 

These assumptions were modified by \citet{lissauer09}
 to take into account the results of new
three-dimensional numerical simulations of protoplanetary disks
accreting onto protoplanets. These simulations employed the  code developed 
by \citet{dangelo03}.  The results from
the 3-D simulations showed that $k_2 \approx 0.25$, that is, the planet can collect material
from the disk only within a quarter of its Hill radius, at best.  The gas accretion rates
from the 3-D simulations, which included the effects of gap opening,
 were used whenever they were less than the
accretion rate needed to maintain the boundary of the planet at $R_A$; 
thus the arbitrary cap on this rate was removed. The parameters
investigated in this set of simulations included (1) the size of the
region from which the protoplanet can accrete, (2) the viscosity in
the protoplanetary disk, which has a strong influence on the disk
accretion rate onto the protoplanet once the envelope has shrunk and the 
planet is massive enough to begin clearing gas from nearby its orbit,
 and (3)  the time-dependence
of the gas density in the protoplanetary disk in the vicinity of the
protoplanet. The grain opacity was set to 2\% interstellar and
$\sigma=10$ g~cm$^{-2}$. The results of these improved simulations
showed that Jupiter's formation time was still less than 3 Myr, and
that a relatively low-viscosity disk ($\alpha \approx 4 \times 10^{-4}$)
is consistent with the formation of planets of about 1 Jovian mass,
while higher viscosity produced planets that were more massive than
most observed extrasolar giant planets. 

In all of the simulations just mentioned, as well as in detailed formation
simulations
by other groups \citep{alibert05,fortier07,ikoma00},
the opacities arising from grains in the protostellar envelope have
been approximated either by interstellar grain values \citep{pollack85}
or by a reduction of those values by some arbitrary amount, typically
a factor 50. However, the simulations of
\citet{podolak03} and \citet{movshovitz08}, in which grain coagulation and sedimentation were calculated
for a few particular protoplanetary envelopes, showed that the ratio
of actual grain opacity to interstellar grain opacity was highly
variable, depending on the depth in the envelope.  Near the surface
of the planet the opacities were close to interstellar, while deep in
the envelope they were reduced by a factor of up to 1000. 
 Those  opacity calculations, 
however, were  not coupled to the evolution of the protoplanet, and they
included only the outer radiative layer in the atmosphere, which typically
has a temperature of 600 K at its inner boundary.

In the present paper we extend the basic grain settling and coagulation 
model of  \citet{movshovitz08} 
by including  the entire depth of the envelope down to a 
temperature of 1800 K, where practically all grains have evaporated. 
Since some layers of the outer part of the envelope are convective, we also modify the
grain simulation to take into account convective effects. We then
couple the grain calculation to the evolution of the protoplanet, 
recalculating the grain size distribution and opacity in every layer
of the protoplanet at every time step of the evolutionary calculation.
These significant improvements in the core accretion model allow us
to investigate the following questions: (1) How is the formation time
for a Jupiter-mass planet  at 5.2 AU modified compared with the earlier 
approximate calculations regarding the opacity? (2)  Most of our previous
simulations for the formation of Jupiter required an initial solid
surface density of 10 g cm$^{-2}$ in the protoplanetary disk. How far can this
assumed density be reduced such that Jupiter still forms within a 
dissipation time of the disk -- roughly 3 Myr -- and what is the resulting
solid core mass? The following sections describe the nature of the grain
simulations, the basic idea of the core-nucleated  accretion model, 
how the two calculations are integrated, and our new results for four
formation runs, up to the onset of rapid gas accretion,
 for a Jupiter-mass planet at 5.2 AU.

\section{Grain Physics\label{sect:grain}}
This section describes, for a protoplanetary envelope with a given
distribution of temperature, density, and convective velocity as a function of distance
to the center,
how grain settling and coagulation modify the grain size distribution and
opacity at each level.
The size distribution of the grains as a function of height is determined
 as follows:  At each atmospheric level the number of grains per unit
volume, $n(m,r,t)$, evolves according to 
\begin{equation}
\frac{\partial n(m,r,t)}{\partial t}=Q(m,r,t)+\left
 [\frac{\partial n(m,r,t)}{\partial t} \right]_{coag} +\left
 [\frac{\partial n(m,r,t)}{\partial t} \right]_{trans}, 
\label{evol}
\end{equation}
where the derivatives are Eulerian.
Here $m$ is the mass of the grain, $r$ is the distance from the center of the planet, and $t$ is the time.

The first term on the RHS of Eq.~(\ref{evol}) is a source term, and refers
to those grains that are carried into the envelope either by the accreted
 gas or by planetesimal ablation.  We do not consider the effects of
 evaporation and condensation on the grain size distribution.   
A grain is instantaneously evaporated when its temperature reaches 
1800 K. (We assume that the grains are composed of silicates. The error introduced by this assumption is probably small compared with the overall uncertainty in the
physics of grain structure and agglomeration.
 This point is furthered discussed below.)

The second term on the RHS of Eq.~(\ref{evol}) refers to the evolution of the
 grain size distribution by coagulation.  We treat this by solving the
 Smoluchowski equation
$$
\left [\frac{\partial n(m,r,t)}{\partial t} \right]_{coag}=\frac{1}{2}
\int_0^mK(m',m-m',r,t)n(m',r,t)n(m-m',r,t)dm'
$$
 \begin{equation}\label{smol}
-\int_0^{\infty}K(m,m',r,t)n(m,r,t)n(m',r,t)dm'.
\end{equation}
The coagulation kernel, $K(m,m',r,t)$ is essentially the probability of a
 collision between grains of mass $m$ and $m'$ that results in the two
 grains sticking together to form a new grain.  The details of how $K$ 
is computed are given in  \citet{podolak03} and \citet{movshovitz08}. 

In practice, the grain size distribution is approximated by a finite number
 of size bins, where the $i$th bin represents a grain of radius
 $$a_i=2^{i/3}a_0$$ and $a_0$ is a scaling constant, set in this case
to 1 $\mu$m. It has been shown \citep{movshovitz08}
that if the minimum grain size in a protoplanetary envelope is 
reduced to 0.1 $\mu$m, the results are practically identical to those 
for a minimum size of 1 $\mu$m, because of rapid coagulation of the
small particles.  The number of bins is
 chosen so that the size range of interest can be adequately represented. 
As a result, the integrals in Eq.~(\ref{smol}) are replaced by sums.

The third term on the RHS of Eq.~(\ref{evol}) is a transport term.  In the absence of convection, the transport is given by
\begin{equation} \label{trans}
\left [\frac{\partial n(m,r,t)}{\partial t}
 \right]_{trans}=-\nabla\cdotp\left[n(m,r,t){\bf v}_{sed}(m,r,t)\right].
\end{equation}
Here ${\bf v}_{sed}$ is the sedimentation velocity of the grain, and it is a
 function of the grain size and the local values of gravity, gas density, and 
temperature \citep{movshovitz08}.

If convection is present, it affects the results in  two ways. 
 First, small grains that are entrapped by a convective eddy reach 
higher speeds than normal, while large grains are less affected by 
these eddies.  This effect causes  the relative velocities between small and large
 grains, and the resultant coagulation kernels, to  change.
  \citet{weidenschilling84} gives expressions for grain relative velocities in
 the limits where both grains are large, one is large and one is small,
 and both small (his Eqs.~15, 16, and 18, respectively). More recently,
\citet{ormel07}  give an expression for the relative velocity between
 particles of arbitrary size (their Eq.~16), that matches the
 \citeauthor{weidenschilling84} expressions in the relevant limits. We use
 the \citeauthor{ormel07}  expression, modified to include the effects of the
 systematic (sedimentation) velocity of the grains. The sedimentation velocity
 shifts the boundary between small (coupled to the gas) and large (uncoupled)
 grains, and the \emph{difference} in sedimentation velocity between two 
grains is also added in quadrature to the relative velocity due to turbulence.

In practice, however, the enhancement of collision probability due to
 turbulence, although fully included in the calculations,
 is not terribly important. For the most part, relative velocities
 between grains of different sizes are determined by the difference in
 sedimentation speed, and the relative velocities between small particles of
 similar size are determined only by their random thermal motion, because
they are coupled to the gas.
 The turbulence
 does add collisions between large grains of equal size. These grains are too
 large for significant thermal motions, and with no difference in
 sedimentation speed they have no appreciable relative velocities, except
 those induced by turbulence. Because this enhancement is only apparent for 
large grains ($\approx 3 \times 10^{-2} - 1$ cm), the effect on the size distribution and opacity is negligible.

A second effect of convection is that the transport term becomes more
 complicated.  \citet{podolak03} treated convective transport with an eddy
 diffusion approximation.  This leads to numerical difficulties and requires
 very short time steps.  As a result, it is unsuitable for use in the current
 evolutionary models.  We have therefore treated convective transport with a
 simplified algorithm that retains the major characteristics of convective
 mixing, explicitly conserves mass, and is very computationally efficient. 
 For each layer in the envelope and for each grain size, we decide if
 convection is important or not based on whether two conditions are satisfied.
The first is that the time it takes for a grain to become entrapped in the
 largest eddy be shorter than the lifetime of that eddy.  If the grain is small
 enough so that the time to entrap it is short, the grain responds to the
 convective motion.  For this calculation, we have taken the length of the
 largest eddy to be 1.5 times the pressure scale height.  The eddy lifetime
 was taken to be this length divided by the speed of the eddy.
The second condition that is relevant to convective transport is that the
 convective speed of the eddy be greater than the sedimentation speed of the
 grain.  If this condition is not met, the grain sediments faster than the
 convection can mix it.  Going back to the first condition, note that
there are actually two requirements for a grain to be carried by an eddy.
First, as stated above, the stopping time must be shorter than the eddy
turnover time. Second, the  stopping time must be 
shorter than the grain crossing time (eddy size/grain velocity). In our
case, the crossing time is $\sim L/v_{sed}$, and the turnover time is
$\sim L/v_{conv}$, where $L$ is the eddy size. Since the second condition requires $v_{conv} >
v_{sed}$, the crossing time is longer than the turnover time, which
we have already required to be longer than the stopping time. Thus
the requirement (stopping time $<$ crossing time) is automatically met.

For each layer in the envelope and for each bin size
 we determine whether or not these two criteria are met.  If they are, convection
 will be the dominant mode of transport.  We then determine, for each grain 
size, which regions (i.e., sets of contiguous layers) of the envelope will mix
 that grain size.  In those regions we determine the total number of grains,
 and redistribute them so that the ratio of mass density of grains to mass
 density of gas is constant. Grain sedimentation velocities range from
 millimeters to meters per second. We find that almost all grains are mixed 
for convective velocities
 larger than $\sim{1000}$ cm s$^{-1}$.

This algorithm has the disadvantage that it allows for either complete
 convective mixing or no convective mixing.  Partial convective mixing is not 
considered.  However, it has the advantage that it explicitly conserves the
 grain mass in the region, and it allows much larger time steps than the eddy
 diffusion algorithm.  Numerical tests show that the resultant grain size
 distributions behave in accordance with what is expected for convective
 mixing, so this seems to be a useful approximation to convective transport.

Once the grain size distribution is established, the opacity is computed
 using an approximation to Mie scattering suggested by Cuzzi \citep[personal communication;][]{movshovitz08}.  The effective cross-sections for
absorption and scattering are calculated for each grain size and
frequency, and combined to produce the opacity $\kappa_\nu (a_i)$. 
This function is then integrated over the grain size distribution and
over the blackbody spectrum to produce the Rosseland mean opacity for
each level in the atmosphere.

We have assumed that the flux of grains into the uppermost layer of the
 envelope is 1\% of the gas accretion mass flux, given a near-solar composition in the protoplanetary disk. This will cause us to
 overestimate the opacity for the
following reason. It assumes that all of the condensed material in this
 region is in the form
of small grains. This is certainly not true, since many of the grains will 
have accreted into
larger planetesimals. These planetesimals are eventually accumulated by the planet, but 
 their grains are released into the envelope at much deeper
levels and thus have much less of an influence on the opacity of
the outer layers.

Our model also assumes that all of the grains are composed of silicates. In
 fact, they
are likely to be composed of a combination of silicates, organic material and
 water, in roughly equal fractions by mass. This has two major consequences. First, although the silicates and 
organics will remain solid at the
temperatures relevant to the outer envelope, the water will evaporate. This
 means that we
have overestimated the mass of grain material at all levels hotter than
$\sim  (170 - 200)$ K and
the true grain opacity will be correspondingly lower.

A second effect is that the bulk density of the grains in these upper layers
 will be lower than we have assumed, and this means that the grains will
 sediment more slowly and their
number density will be higher than we have computed. As a result there should
 be some increase in the opacity. Below we estimate the
 magnitude of this effect.

    In the uppermost layers of the envelope, there is very little grain growth
 and most of the grains are the same size \citep{movshovitz08}.
 To first approximation, the
time for such a grain to settle out of a layer of thickness $l$ is
\begin{equation}
t_{sed} = l/v_{sed}
\end{equation}
where $v_{sed}$ is the sedimentation speed. The time between grain
collisions is roughly
\begin{equation}
t_{coag} = \frac{\Lambda}{v_d}.
\end{equation}
Here $\Lambda$ is the mean free path between grains, and $v_d$ is the
thermal speed of a grain.

If $t_{coag}$ is short compared to $t_{sed} $, the grains will collide and grow
before they sediment out of the region. The grain size can be estimated
by setting $t_{sed}  \approx t_{coag} $. 
The mean free path between grain collisions will be 
\begin{equation}
\Lambda = \frac{1}{\sqrt{2} (2a)^2n},
\nonumber
\end{equation}
where $n$ is the number density of grains, and $a$ is the grain radius.
 The Knudsen number in
this region is large so that the sedimentation speed can be written as
\begin{equation}
v_{sed} = \frac{\rho_d}{\rho_gv_g} ga, 
\nonumber
\end{equation}
where $g$ is the acceleration of gravity, $\rho_d$ is the bulk density of
a single grain, $\rho_g$ is the mass density of the  gas,
and $v_g$ is the thermal speed of a gas molecule. The mass flux through the
layer is denoted by  $F$, where 
\begin{equation}
F=\frac{4 \pi}{3} n \rho_d v_{sed} a^3.
\end{equation}
The thermal speed of a molecule (or grain) of mass $m$
 is given by:
\begin{equation}
v = \sqrt{\frac{8kT}{\pi m}},
\nonumber
\end{equation}
where $T$ is the temperature and $k$ is Boltzmann's constant.
Setting $t_{sed} =  t_{coag}$ and using the above relations, we find
\begin{equation}
a = \left[\frac{3\sqrt{12kT} F l \rho_g^2v_g^2}
{\pi \rho_d^{7/2} g^2}\right]^{2/9}.
\end{equation}
The cross-section for scattering or absorption is the geometric cross-section,
$\pi a^2$,  multiplied
by some efficiency factor, $Q$, that must be computed from Mie theory. 
Typically, $ Q <<   1$ for
grains much smaller than the wavelength of the impinging radiation, and
 roughly equal to
one for larger grains. Since we are interested in Rosseland mean opacities at
 temperatures of around 100 K, the wavelengths will be in the $20 - 30$ micron
 range and the grains are in
the 1 micron range in the upper levels of the envelope. The corresponding
 size parameter is
\begin{equation}
x = \frac{2 \pi a}{\lambda}, 
\nonumber
\end{equation}
where $\lambda$ is the wavelength of the radiation.

For small $x$ the extinction is mostly due to absorption, and a useful
 approximation for
$Q$ is \citep{movshovitz08}:
\begin{equation}
Q = \frac{24 x n_rn_i}{(n_r^2 + 1)^2},
\nonumber
\end{equation}
where $n_r$ and $n_i$ are the real and imaginary indexes of refraction,
 respectively. Typically
$n_r \approx  1.5 $ and $n_i \approx 0.1$ so $Q \approx 0.2x$. 
In our case, $x \approx 0.25 $, so $Q \approx 0.05$.

The opacity per gram of gas is      
\begin{equation}
\Sigma = \frac{Q \pi a^2n}{\rho} = \frac{3QFv_g}{4\rho_d^2ga^2}.
\label{eq:sig}
\end{equation}
All of these quantities are interconnected, since the grain size $a$ 
is also a function of grain
density to the power $-(7/9)$ . As a result, the opacity in these upper
 layers is expected to
depend only weakly on the assumed grain density. Thus, even if the grains are
 made of pure
silicates with a density of $\approx  3$  g cm$^{-3}$ or of pure ice with a
 density 1/3 as large,    the opacity
should change by roughly a factor of 1.6. In practice the change should be
 even smaller.


%

\section{Core Accretion Model\label{sect:core}}
The physics of the core--nucleated accretion model are for the most part 
described by \citet{pollack96}, \citet{bodenheimer00}, 
\citet{hubickyj05}, and \citet{lissauer09}.
The calculation employs the three basic elements mentioned
in Section \ref{sect:intro}. The first element is essentially
Eq.~(\ref{eq:saf}), which is used to calculate the accretion
rate of solids. The value of  $F_g$ is obtained from the numerical fits of
\citet{greenzweig92}, assuming a planetesimal radius of
100 km. Other more detailed calculations of the core accretion
rate \citep{kokubo00} show in general that the rates calculated
by \citet{greenzweig92} should be reduced because they
do not completely take into account the stirring and increase in
relative velocity $v$ of the planetesimals as a result of gravitational
interactions with the embryo. However, the
accretion rate for smaller planetesimals is 
faster than that for 100 km. Thus our accretion rate should be a
reasonable approximation to the actual rate assuming that the 
planetesimals determining the velocity distribution are somewhat
smaller than 100 km.

As in previous calculations, we take the feeding zone from which the
embryo can accrete planetesimals to extend 4 Hill sphere radii
on either side of the orbit, and assume that the surface density $\sigma$
is spatially
uniform within that zone. As the planet accretes material,
$\sigma$ in the feeding zone is recalculated, again assuming
uniform density, taking into account depletion of planetesimals
by accretion onto the embryo and expansion of the feeding zone  into
undepleted regions as the embryo's mass increases. 

The second element of the code calculates the  interactions
between planetesimals and the gaseous envelope of the protoplanet
as they fall through it \citep{podolak88}. The equations of
motion of the planetesimals are integrated, for a variety of
assumed impact parameters, taking into account the gravitational
fields of the core and the envelope as well as gas drag. The
critical impact parameter inside of which the planetesimal is
captured is determined, and the effective capture radius $R_{\rm core}$
(Eq.~\ref{eq:saf}) is assumed to be the periapsis altitude of
the critical orbit.
This radius can be several times larger than the actual solid core
radius, once a significant gaseous envelope has accreted, so this effect
is important in speeding up the accretion rate of solids. 

For those planetesimals with impact parameters less than the 
critical value, trajectories are calculated to determine the 
deposition of mass and energy into the envelope. Effects included are
ablation and evaporation of planetesimal material, heat associated
with phase changes, and fragmentation when the dynamical pressure
of the gas exceeds the compressional strength of the planetesimal.
During the earliest phases, when the envelope has low mass, 
planetesimals plunge all the way to the core. At later stages they
can be fully ablated or fragmented in the envelope. It is assumed
that the remains of the planetesimal later sink to the core,
releasing gravitational energy in the process \citep{pollack96}.
This assumption is not entirely accurate: 
\citet{Iaroslavitz07} have made a calculation of how
much of the heavy-element material should actually dissolve in
the envelope. Ice should be almost entirely dissolved, but most of the
organics and rock would sink to the core or to the layers just outside it.
They conclude that the overall evolution would  not be substantially
changed if this effect were included;  
the accretion of the protoplanet would be slightly
sped up and the core mass would be reduced to about 2/3 the calculated
value, with little
change in the overall metal content. 

The third element of the simulation
is the solution of the four differential equations
of stellar structure for the gaseous envelope, with the nuclear
energy term replaced by the energy delivered by accreting planetesimals
and with the gravitational energy terms included. The adiabatic
temperature gradient is assumed in convection zones. At the core-envelope
interface the radius is set to that of the outer edge of the core,
calculated from $M_{\rm core}$ and the assumed core mean density of
3.2 g cm$^{-3}$. Also, the luminosity $L_r$ is set to the energy
deposition rate for the planetesimals that hit the core.
Outside the core, the energy supplied by ablated and fragmented
planetesimals is included as a source term in the energy equation,
along with the energy derived from gas accretion and compression.

At the surface, mass is added at a sufficient rate to maintain
an outer radius $R_p = R_A$, and Eq.~(\ref{eq:ra}) is used with
$k_2=0.25$, as reported in \citet{lissauer09} from the results of
three-dimensional calculations of disk-planet interaction.
Otherwise the density and temperature at the surface are set to
assumed nebular values $\rho_{\rm neb}$, $T_{\rm neb}$, respectively.
These values are constant in time. The  simulations are stopped when
the gas accretion rate reaches the disk flow limit  d$M_{\rm lim}$/d$t$, 
as  determined from 
three-dimensional calculations with disk
viscosity parameter $\alpha=4 \times 10^{-4}$ \citep{lissauer09}.
The lower curve in
figure~3 of that paper gives the limiting gas accretion rate as a 
function of planet mass. In  the various cases of the present calculation,
 d$M_{\rm lim}$/d$t \approx (1.5~{\rm to}~3) \times 10^{-2}$
 M$_\oplus$ yr$^{-1}$ when the limit is first reached. 

The equation of state of the gas is taken from \citet{saumon95},
interpolated to our assumed composition of hydrogen mass fraction
X~=~0.74, helium mass fraction Y~=~0.243, and metal mass fraction 
${\rm Z}=1 - {\rm X} - {\rm Y} = 0.017$. Although the equation of state in the outer, low-density
layers is essentially that of an ideal gas, the inner regions near the
core can be significantly non-ideal once the envelope has acquired
sufficient mass.

 The Rosseland mean opacity calculation has three 
components. At temperatures above 3500 K the molecular opacities of
\citet{alexander94} are used. In practice the details of the
opacities in this region are unimportant because the energy
transport is almost always by convection. In the temperature range
200--3500 K the molecular opacities, without grains, of \citet{freedman08} are used. The grain opacity calculation described in 
section \ref{sect:grain} is then included in the temperature range
100--1800 K. In the higher end of this range the grain opacity can
be so low at times that the molecular opacity dominates. Grains are
assumed to be completely evaporated above 1800 K.

The grain calculation and the stellar envelope calculation are
combined as follows. After each evolutionary time step $\Delta t$ for
the envelope, grains are added at the surface in accordance with
d$M_{\rm env}$/d$t$, assuming a dust-to-gas ratio of 0.01. Grains are
also added where the ablation/fragmentation calculation for the
incoming planetesimals deposits them, typically fairly deep in the
envelope. For the grains, 34 size bins are used, with the smallest
at $1.26$ $\mu$m and the largest at 2.58 mm. Incoming grains, represented
by the term $Q(m,r,t)$ in Eq.~(\ref{evol}), are always placed
in the smallest bin. The envelope calculation then provides to the
grain calculation a table of temperature, density, and convective
velocity as a function of radius, as well as the core and envelope
masses.  Velocities are estimated by
the stellar structure code through the mixing-length theory. An
arbitrary number of distinct convection zones is  allowed.

The grain evolution, involving settling and coagulation, then 
proceeds in the region from the surface down to 1800 K. A large
number of sub-timesteps is used, adding up to a total time of $\Delta t$,
as the number of grains in any size bin is not allowed to change
by more than 4\% per sub-timestep. The number of sub-timesteps can be
10,000 to 100,000, so this part of the calculation is quite time-consuming.
Once the grain calculation has been completed, with a new opacity
provided at each depth, the opacities are fed back into the envelope
calculation in the form of a table of Rosseland mean opacity as a
function of mass fraction, integrated inward from the surface. Mass
fraction is used rather than temperature or density because, first,
the temperature is practically isothermal in the outer layers, and, second,
temperature and density values can fluctuate somewhat as a result of
the numerical procedure for adding mass to the envelope. The current
grain size distribution is stored at each depth for use in the next
time step. The envelope calculation then proceeds to the next model.
Once the model has been completed, the Lagrangian zones have moved in
radius with respect to the previous model. The grain number densities, 
tabulated as a function of radius, are interpolated onto the new grid
by a procedure that conserves the total number and mass of grains.
The results for the grain opacity are qualitatively quite similar to
those obtained for specific envelope models by \citet{movshovitz08}. In the outer zones, where small grains are continuously
added from the protoplanetary disk, the opacity generally falls in the
range 0.1--1 cm$^2$ g$^{-1}$, not too different from opacities
derived from grains with interstellar size distribution and 
abundance. In contrast, at
the deeper levels the typical grain size increases sufficiently so that
the mean opacity can drop by up to      4 orders of magnitude relative
to interstellar values!

\section{Calculations and Results}

The planet initially consists of a core of 1 M$_\oplus$ and an
envelope of about $10^{-5}$ M$_\oplus$. The protoplanet is located at 5.2 AU in a solar nebula disk, with
the solid surface density $\sigma$ treated as a parameter. Based on our past calculations for the time required to reach $1\;M_\oplus$
\citep{hubickyj05,lissauer09}, the starting time is set
to $2.8\times{10^5}$ yr, $4.85\times{10^5}$ yr, and $7.28\times{10^5}$ yr
for $\sigma=10$, 6, and 4 g cm$^{-2}$, respectively. No orbital
migration is included. The quantity $T_{\rm neb}$ is set to 115 K, and
$\rho_{\rm neb}=\sigma_{XY}/(2H)$, where $\sigma_{XY}=70 \sigma$.
The scale height of the gas in the disk  $H=0.05 a$, where $a$ is the orbital distance
from the star. Once started, the evolution usually consists of three main
phases. The first involves primarily accretion of solids onto the core, with a
relatively low-mass envelope and a low gas accretion rate. The solids
accretion rate slows down significantly near the point where the isolation
mass  ($M_{\rm iso}$) for the core is reached; for $\sigma=10$ at 5.2 AU this mass is
about 11.5 M$_\oplus$. During the second phase, the gas accretion rate is
about 3 times as high as the core accretion rate, and both are nearly
constant in time. The envelope mass builds up relative to the core mass,
which grows     slowly. The third phase, rapid gas accretion, starts
just prior to crossover mass, $M_{\rm core} = M_{\rm env} 
= \sqrt{2} M_{\rm iso}$. The calculations reported here stop at a point
well beyond crossover, where d$M_{\rm env}$/dt = d$M_{\rm lim}$/dt. Beyond
that point, accretion up to $\approx 1$ M$_{\rm Jup}$ is very fast,
and the core mass does not change appreciably.  Examples of this final
phase are given in \citet{lissauer09}; note especially  Cases 2l and 2lJ,
 calculated with $\sigma=10$  g cm$^{-2}$.

Three main cases were calculated, with $\sigma = $ 10, 6, and 4 g 
cm$^{-2}$ and with all other parameters identical.  These will be referred
to as runs $\sigma$10, $\sigma$6, and $\sigma$4, respectively.
 A fourth case ($\sigma$10R) was
calculated with $\sigma = $ 10 g cm$^{-2}$ but without convective
effects included in the grain calculation, although they were
included in the actual planetary model calculation. The most closely
analogous  comparison case from \citet{lissauer09} was their Case 2lJ
with $\sigma = $ 10 g cm$^{-2}$ and the same outer boundary conditions
as used here; however \citeauthor{lissauer09}\ simply set the grain opacity equal
 to 2\% of interstellar values. That calculation reached a crossover mass
of 16.16 M$_\oplus$ at a time of 2.3 Myr, and a final total
mass of 1 M$_{\rm Jup}$ at 2.6 Myr. The input parameters of the four
runs are shown in Table~1, which includes $\sigma$, the gas
surface density $\sigma_{XY}$,  the surface boundary temperature 
$T_{\rm neb}$, a column ``conv?", which indicates whether convection
was included in the grain calculation or not, and the isolation mass
$M_{\rm iso}$.

\begin{table}[tbp]
 \caption{Input Parameters}\label{table:1}
 \centering
 \vspace*{1ex}%
 \resizebox{1.0\linewidth}{!}{%
 \small
 \begin{tabular}{|l||ccccc|}
 \hline
 Run    
 & $\sigma$ (g/cm$^2$)                         & $\sigma_{XY}$  (g/cm$^2$)  
 & $T_{\mathrm{neb}}$ (K)  &  conv?     
 & ${M}_{\rm iso}$ (M$_{\oplus}$)   
   \\
 \hline\hline
 $\sigma$10    
 &         10   & 700                       
 &  115                    & Y          
 &  $11.56$    
  \\
 \hline
 $\sigma$10R   
 &         10                  & 700    
 &  115                    &  N
 &  $11.56$                 
    \\
 \hline
 $\sigma$6      
 &     6                      & 420    
 &  115                    &  Y           
 &  $5.37 $     
   \\
 \hline
 $\sigma$4          
 &     4   & 280    
 &  115                    & Y         
 &  $2.92$ \\                 
 \hline

 \end{tabular}
       }
\end{table}

\begin{figure}[tp]
\includegraphics[width=\linewidth,keepaspectratio,clip]{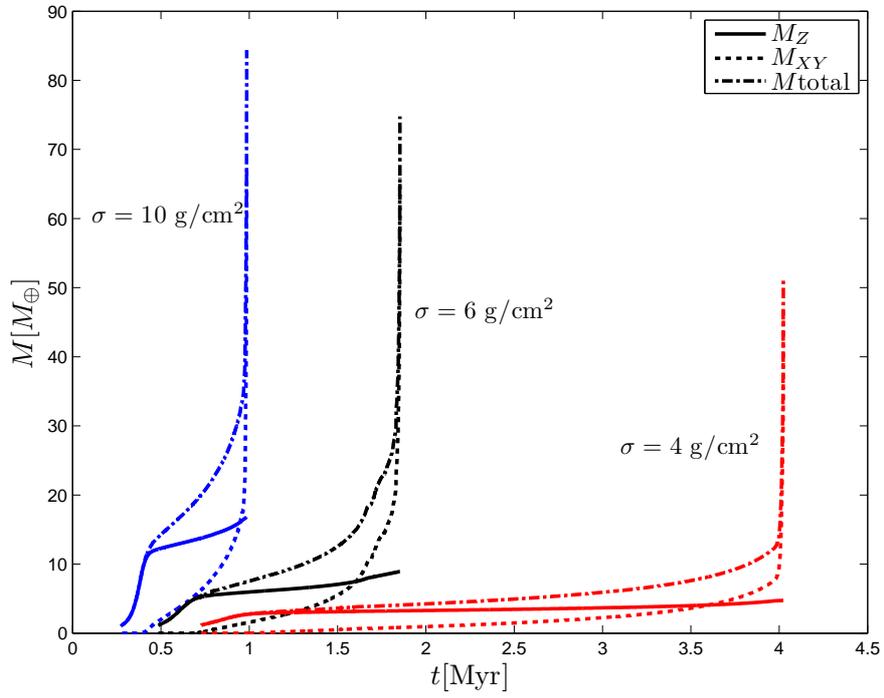}
\caption{\label{fig:f1}Mass of the protoplanet as a function of time for three cases
that include grain settling and coagulation. The solid lines denote the
mass of the core, the dotted lines the mass of the H/He envelope, and
the dot-dashed lines the total mass.  All cases plotted include 
convection in the grain calculations.  The assumed solid surface
density $\sigma$ is indicated for each set of curves.
}
\end{figure}
\begin{figure}[tp]
\includegraphics[width=\linewidth,keepaspectratio,clip]{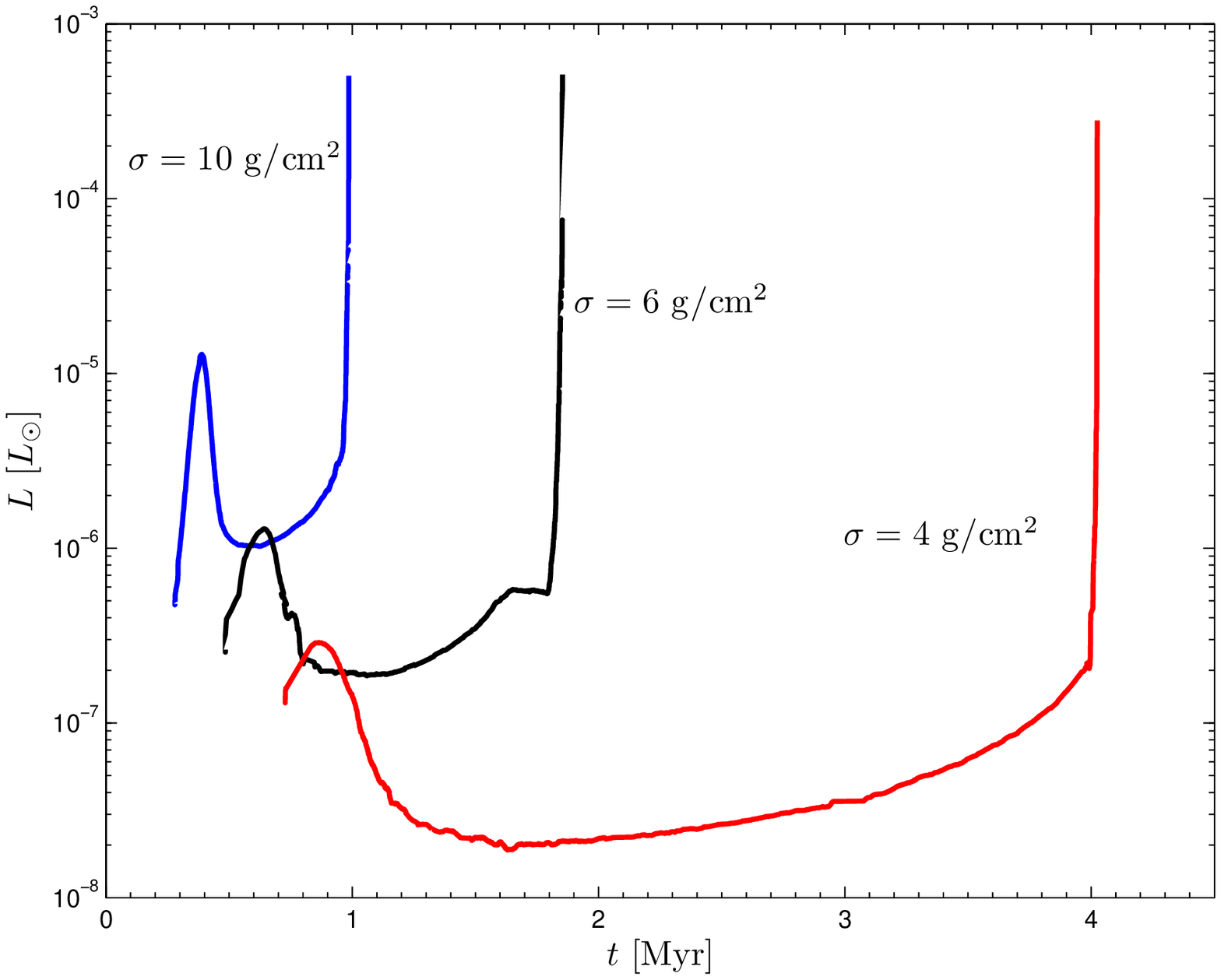}
\caption{\label{fig:f2}The protoplanet's luminosity as a function of time for the 
same cases as shown in Fig. \ref{fig:f1}.
}
\end{figure}

The results of our calculations  with convection
 are shown in Figs.~\ref{fig:f1} and
\ref{fig:f2}. For the $\sigma=10$ case, the formation of
the planet is significantly speeded up in comparison to case 2lJ.
The early core accretion phase is almost identical to that in case 2lJ
as the envelope opacity has little effect on the rate of buildup of
the core. The first luminosity peak is at a level of $10^{-5}$ L$_\odot$
at 0.4 Myr;  the isolation mass is approached at about 0.45 Myr.
However Phase 2, during which core and envelope accretion rates are
relatively low, is significantly shorter in the present calculation.
The main effect of the grain settling calculation is to reduce the
average opacity in the envelope, allowing it to lose heat more
readily, resulting in a higher rate of envelope contraction and
therefore needing a higher gas accretion rate to keep the outer
boundary at $R_A$. The crossover mass is almost identical to that
in case 2lJ, but it is reached at a time of only 0.97 Myr. Beyond that
point the gas accretion rate increases rapidly, reaching the 
limiting rate of $2.7 \times 10^{-2}$ M$_\oplus$ yr $^{-1}$ at 
$1.03 \times 10^6$ yr with $M_{\rm core} = 16.8$ M$_\oplus$
and $M_{\rm env} = 56.8$ M$_\oplus$. By way of comparison, case
2lJ reached the same d$M_{\rm lim}$/d$t$  with  $M_{\rm core} = 17.0$ M$_\oplus$
and  $M_{\rm env} = 55.5$ M$_\oplus$, but the time was $2.45 \times 10^6$ yr.

\begin{figure}[tp]
\includegraphics[width=\linewidth,keepaspectratio,clip]{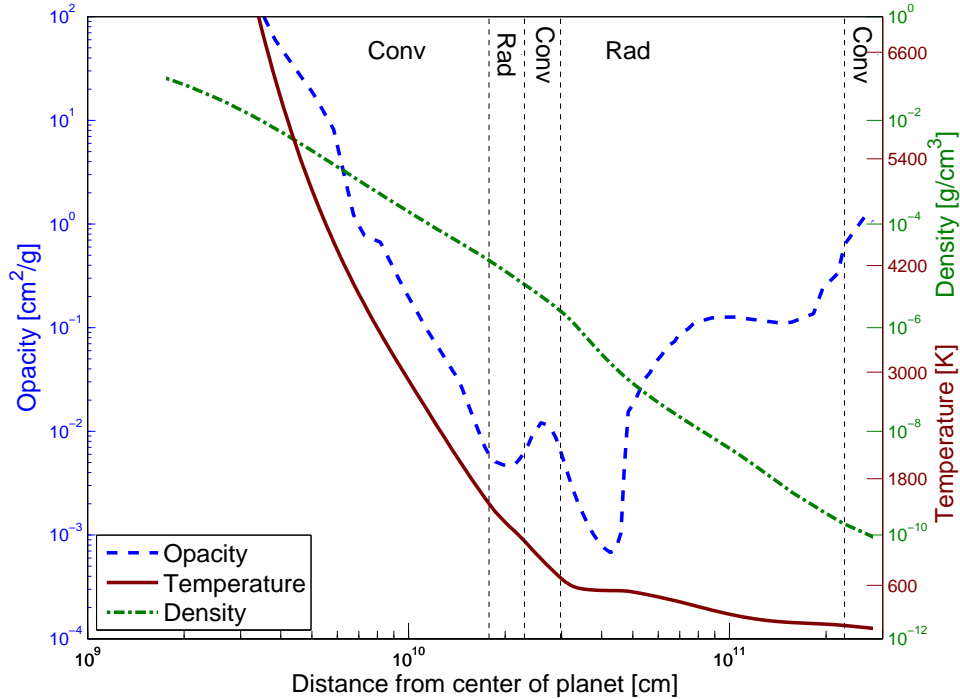}
\caption{\label{fig:f3}Structure of the $\sigma$10 case at an age of $4.9 \times 10^5$ yr,
when the core mass is 12.07 M$_\oplus$ and the envelope mass is 1.53
M$_\oplus$. Convective and radiative zones are indicated.            
The total radius is $2.8 \times 10^{11}$ cm.
The temperature at the core/envelope interface is 1.33 $\times 10^4$ K.
}
\end{figure}

\begin{figure}[tp]
\includegraphics[width=\linewidth,keepaspectratio,clip]{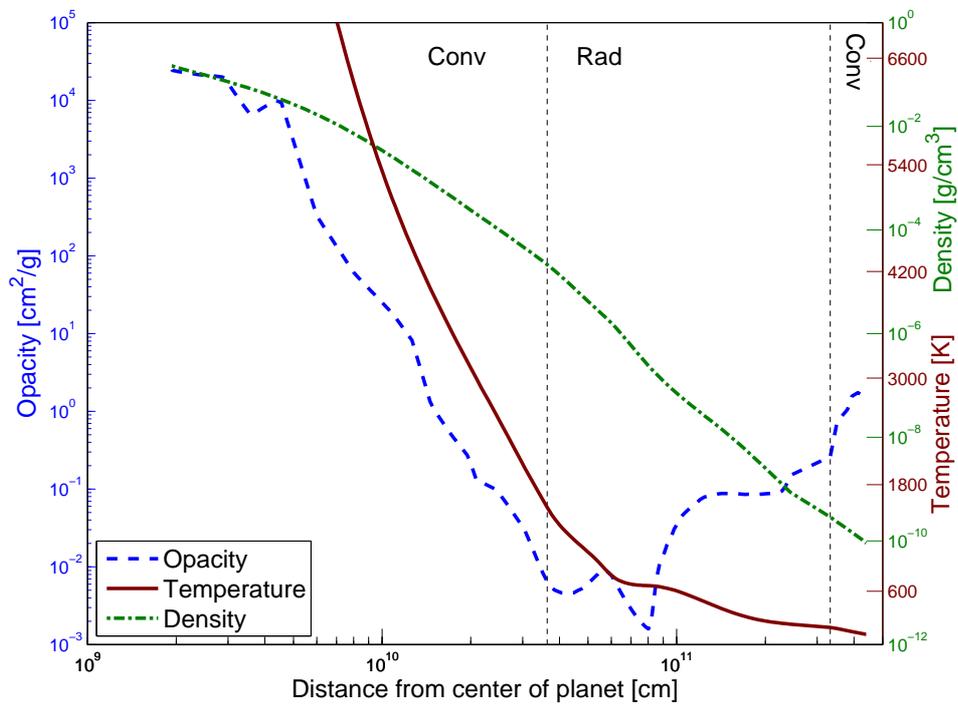}
\caption{\label{fig:f4}Structure of the $\sigma$10 case at crossover, age $9.68 \times 10^5$ 
yr with core mass 16  M$_\oplus$.  Convective and radiative zones are indicated.
The total radius is $4.19 \times 10^{11}$ cm.
The temperature at the core/envelope interface is 1.86 $\times 10^4$ K.
}
\end{figure}

Thus the main results  are that, as in previous work where we examine the effect of opacity
on planetary formation \citep{pollack96,hubickyj05}, 
the final core mass
is very insensitive even
to large changes in the opacity. Furthermore, compared to calculations
with full interstellar grain opacity, not including grain coagulation
and settling \citep{hubickyj05}, the formation time for
Jupiter is reduced by a factor 6.  The main contributor to this 
result is the sharp reduction in the opacity that appears in the outer
radiative zone at temperatures $\approx$ 500-600 K. This feature is visible
in the structure plots shown in Figs.~\ref{fig:f3} and \ref{fig:f4}.
Interior to this minimum, molecular opacities rather than grain opacities
dominate, and the opacity increases with temperature.

The properties of Runs $\sigma$10, $\sigma$10R, $\sigma$6, and $\sigma$4
 at particular times in the
evolution are summarized in Table 2.  Here $M_Z=M_{\rm core}$ and 
$M_{XY} =  M_{\rm env}$. Case $\sigma$10R is not included in 
the figures, because, as shown in Table 2,  the results are very similar to
those of $\sigma$10. The small differences in the two cases are mainly a result
of numerical fluctuations in the calculations.  The properties are indicated at the following times:
the maximum in luminosity that occurs during the main core accretion phase,
the transition between Phase 1 and Phase 2 where the isolation mass of
the core is approximately reached, a time in the middle of Phase 2, the
time at which the crossover mass ($M_{\rm core} = M_{\rm env}$) is attained,
and the time when the limiting gas accretion rate is reached.
Following this last time, the planet quickly reaches its final mass
through rapid gas accretion; the details of this final phase are
treated in \citet{lissauer09}.

\begin{table}[tbp]
 \caption{Results}\label{table:2}
 \centering
 \vspace*{1ex}%
 \resizebox{0.7\linewidth}{!}{%
 \small
 \begin{tabular}{|l|l||c|c|c|c|}
 \cline{1-1}
 \cline{2-2}
 \cline{3-3}
 \cline{4-4}
 \cline{5-5}
 \cline{6-6}
 \hline
 \cline{1-1}
 \cline{2-2}
 \cline{3-3}
 \cline{4-4}
 \cline{5-5}
 \cline{6-6}
 \hline
 &                                & $\sigma$10  & $\sigma$10R           & $\sigma$6        
 & $\sigma$4                  \\
 \hline\hline
 \cline{1-1}
 \cline{2-2}
 \cline{3-3}
 \cline{4-4}
 \cline{5-5}
 \cline{6-6}
 \hline
 \raisebox{-8.5ex}[+0.0ex][-0.0ex]{\parbox{5.5em}{%
 \textsc{First\\ Luminosity\\ Peak}}}
 & Time                           & 0.408        & 0.405         & 0.654
 &  0.888                 \\
 & $M_{Z}$                        & 8.53         &  8.12         & 3.98
 &  2.12          \\
 & $M_{XY}$                       & 0.064        & 0.044         & 0.0061
 &  0.0014               \\
 & $\dot{M}_Z$                    & $1.3 \times 10^{-4}$ & $1.25 \times 10^{-4}$ & $2.3 \times 10^{-5}$
 & $7.1 \times 10^{-6}$  \\
 & $\log{L}$                      & -4.85        & -4.89         & -5.84
 & -6.53                \\
 & $R_p$                          & 28.8         & 27.7          & 17.3 
 & 9.86                  \\
 \hline
 \raisebox{-8.5ex}[+0.0ex][-0.0ex]{\parbox{5.5em}{%
 \textsc{End of\\ Phase~1}}}
 & Time                           & 0.451       & 0.454          & 0.838
 & 1.19              \\
 & $M_{Z}$                        & 11.5        & 11.5           & 5.60 
 & 2.97                 \\
 & $M_{XY}$                       & 0.67        & 0.61           & 0.63
 & 0.18                  \\ 
 & $\dot{M}_Z$                    & $2.5\times 10^{-5}$& $2.1 \times 10^{-5}$  & $1.3\times 10^{-6}$
 & $4.9\times 10^{-7}$   \\
 & $\log{L}$                      & -5.46       & -5.52          & -6.64
 & -7.49              \\
 & $R_p$                          & 34.7        & 35.2           & 23.1 
 &  15.7                 \\
 \hline
 \raisebox{-8.5ex}[+0.0ex][-0.0ex]{\parbox{5.5em}{%
 \textsc{Mid\\ Phase~2}}}
 & Time                           & 0.775      & 0.781           & 1.295
 & 2.738               \\ 
 & $M_{Z}$                        & 14.0       & 14.0            & 6.50
 & 3.54                 \\ 
 & $M_{XY}$                       & 7.0        & 7.0            & 3.25
 & 1.77                 \\
 & $\dot{M}_Z$                    & $7.9\times 10^{-6}$& $1.2 \times 10^{-5}$  & $2.1 \times 10^{-6}$
 & $4.4 \times 10^{-7}$   \\
 & $\dot{M}_{XY}$                & $2.7\times 10^{-5}$ & $3.8 \times 10^{-5}$ & $7.5\times 10^{-6}$
 & $1.6\times 10^{-6}$ \\
 & $\log{L}$                      & -5.88      & -5.77          & -6.65
 & -7.52                \\
 & $R_p$                          & 47.4       & 48.0           & 30.8  
 & 21.8                 \\
 \hline
 \raisebox{-8.5ex}[+0.0ex][-0.0ex]{\parbox{5.5em}{%
 \textsc{Crossover\\ Point}}}
 & Time                           & 0.968      & 0.947          &  1.625
 & 3.618                \\
 & $M_{\mathrm{cross}}$           & 16.09      & 16.11          & 7.50 
 & 4.09                  \\
 & $\dot{M}_Z$                    & $2.1\times 10^{-5}$ & $1.8 \times 10^{-5}$  & $4.4\times 10^{-6}$ 
 & $9.9\times 10^{-7}$         
                                           \\
 & $\dot{M}_{XY}$        
 & $1.8\times 10^{-4}$ & $6.1 \times 10^{-5}$     & $2.2 \times 10^{-5}$ & $4.2 \times 10^{-6}$

                                           \\
 & $\log{L}$        & -5.54       & -5.52                       & -6.27
 & -7.13            \\
 & $R_p$                          & 57.8       & 60.1           & 38.6 
 & 25.5                 \\
 \hline
 \raisebox{-6.5ex}[+0.0ex][-0.0ex]{\parbox{5.5em}{%
 \textsc{Onset of\\ limited\\ gas\\ accretion}}}
 & Time                           & 1.003      & 0.997          &  1.865 
 & 4.038       \\
& $M_Z$                           & 16.8       & 17.3           &  8.92 
& 4.74             \\
& $M_{XY}$                        & 56.8       & 52.1           & 54.3 
& 34.0             \\
&$\dot{M}_{XY}$                   &$2.7 \times 10^{-2}$ &$2.9 \times 10^{-2}$  & $1.84 \times 
10^{-2}$                    & $1.51 \times 10^{-2}$           \\  
& & & & & \\
 \hline
 \end{tabular}
                              }\\[1ex]
 \hspace{0.2in}%
 \begin{minipage}[t]{\linewidth}
 \footnotesize
 \noindent%
  Time is in units of millions of years, Myr.\\
  Mass ($M_Z$ and $M_{XY}$) is in units of Earth's mass, M$_{\oplus}$.\\
  The accretion rate ($\dot M_Z$ and $\dot M_{XY}$) is in units of Earth masses per 
         year, M$_{\oplus}/$yr.\\
  Luminosity ($L$) is in units of solar luminosity, 
         L$_{\odot}$.\\
  Radius ($R_p$) is in units of Jupiter's present equatorial radius, R$_J$.
 \end{minipage}
\end{table}

Case $\sigma$6 is calculated with $\sigma=6$, and as expected the formation
time is longer, but still less than that for the calculation done by \citet{hubickyj05} for this level of $\sigma$.
The final core mass is, however, significantly reduced, to 8.92 M$_\oplus$.
The first luminosity peak is reached in 0.65 Myr, somewhat
later than in case $\sigma$10. The core accretion rate is of course slower
in case $\sigma$6, but the available planetesimal supply is considerably less,
a compensating effect. The maximum luminosity is typically reached
when $\sigma$ has been reduced  (by accretion onto the planet) 
to 18\% of its initial value.  For case $\sigma$6, the maximum  is a
  factor 10 less than in case $\sigma$10.
Phase 2 is reached at a time of $ 8.4 \times 10^5$ yr with a core
mass of 5.6 M$_\oplus$, half that of case $\sigma$10 at the corresponding epoch.
Mainly because of the reduced core mass \citep[see][]{pollack96}, the
luminosity is considerably lower in Phase 2 than it is for case $\sigma$10,
therefore energy is released less rapidly, contraction is slower, and
the rate of gas accretion is reduced. Thus Phase 2 is lengthened, but
its duration remains less than 1 Myr, giving a total formation time
of 1.86 Myr.

\begin{figure}[tp]
\includegraphics[width=\linewidth,keepaspectratio,clip]{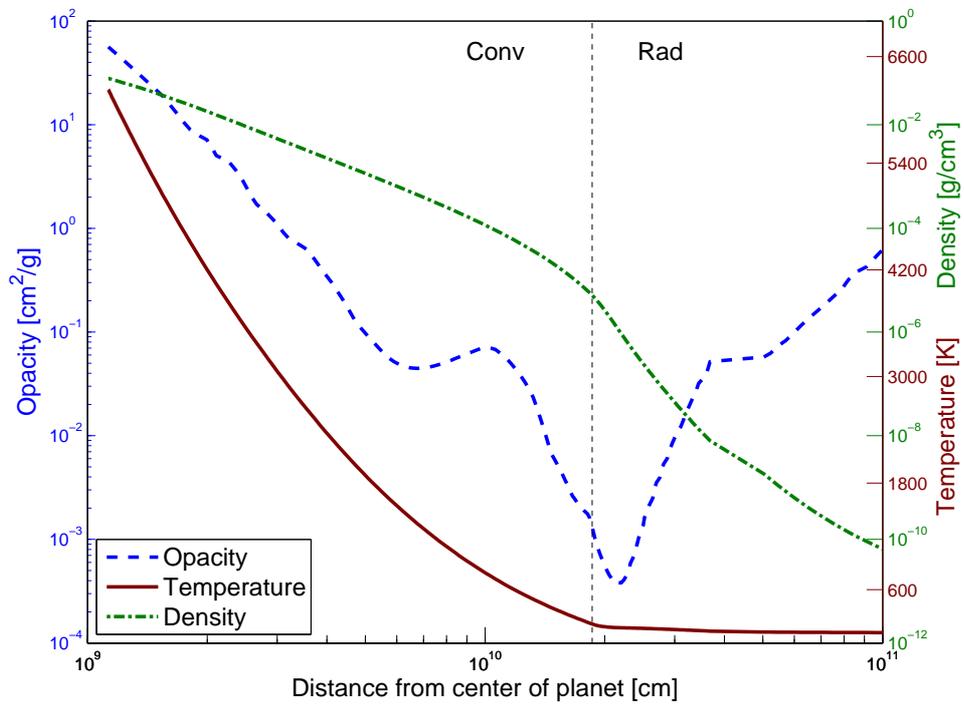}
\caption{\label{fig:f5}Structure of the $\sigma$4 case during phase 2 at an age of
 $1.768 \times 10^6$ yr,
when the core mass is 3.21  M$_\oplus$ and the envelope mass is 0.74
M$_\oplus$.  The convective-radiative boundary is indicated.     
The total radius is $1.20 \times 10^{11}$ cm.
The temperature at the core/envelope interface is 6230 K.
}
\end{figure}

\begin{figure}[tp]
\includegraphics[width=\linewidth,keepaspectratio,clip]{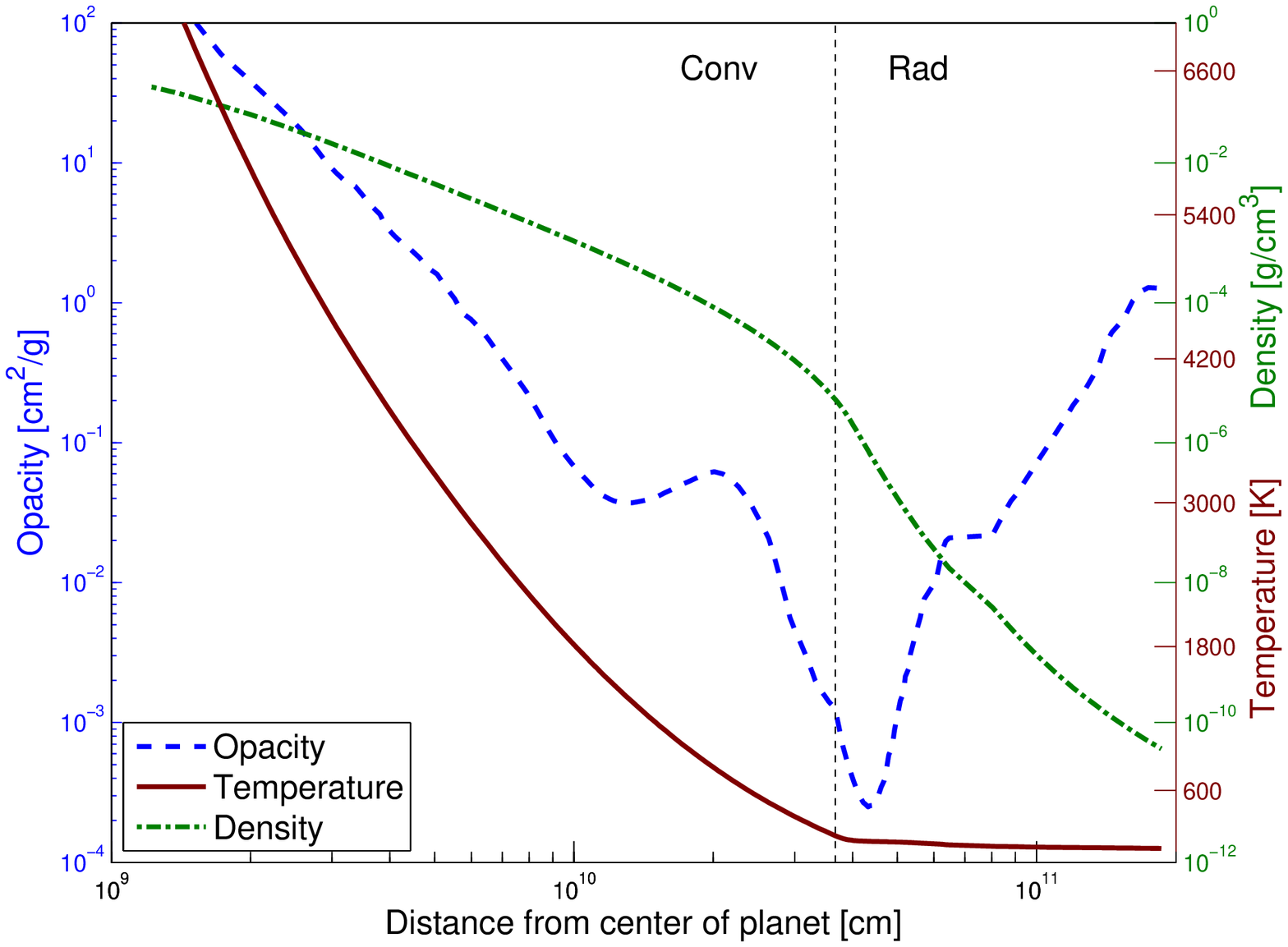}
\caption{\label{fig:f6}Structure of the $\sigma$4 case at crossover, age $3.618 \times 10^6$ 
yr with core mass only 4.1  M$_\oplus$.  The convective-radiative boundary
is indicated. The total radius is $1.93 \times 10^{11}$ cm.
The temperature at the core/envelope interface is 7680 K.
}
\end{figure}

The case $\sigma$4 represents a disk with  $\sigma=4$ g cm$^{-2}$ at 5.2 AU.
This value is only
slightly greater than that of the minimum mass solar nebula, but note
that our planetesimal accretion prescription neglects expulsion of solids
from the planet's accretion zone, so a somewhat higher $\sigma$ is 
probably required in order to reproduce the core accretion rate calculated
in this case. The first
luminosity peak occurs at about the same time as in $\sigma$6, but
the value of the luminosity is a factor of 5 lower. Crossover is reached at
a core mass of only 4 M$_\oplus$ and a luminosity of $10^{-7}$ L$_\odot$.
As shown in Figs.~\ref{fig:f5} and~\ref{fig:f6}, the temperatures at a given density in 
the envelope structure are considerably lower than in case $\sigma$10.
The  planet's internal structure, both in Phase 2 and at crossover, is characterized by an
inner convection zone and an outer radiative zone, with the boundary
at a temperature of only 200 K. The deep minimum in opacity just outside
this boundary  allows for a high rate of envelope radiation and the 
very rapid addition of mass.  The onset of limited gas
accretion is reached after a time of 4.04 Myr, approximately the mean
lifetime as observed for protostellar disks. The heavy element core has a mass
of only 4.74 M$_\oplus$.

\section{Discussion and Conclusions}

The formation phase of Jupiter, as determined by the concurrent accretion
of solid particles and gas, has been calculated under the following major
assumptions: (1) The protoplanet remains fixed  in position at 5.2 AU. 
(2) The planet is assumed to exhibit runaway accretion of solids within a disk
of 100 km planetesimals.
The planetesimals are well mixed
within the feeding zone of the planet; thus their surface density is
constant in space but varies in time as the planet accretes. Migration
of planetesimals into or out of the feeding zone is not
considered. (3) The protoplanet is assumed to be the sole dominant
mass in the region; other embryos which might compete for planetesimals
are not considered.
(4) The solids accretion rate is determined from Safronov's formula (eq.~1)
with gravitational enhancement factor obtained from the 3-body numerical fits
of \citet{greenzweig92}.  (5) All of the planetesimal
material accreted by the protoplanet is assumed to fall into
the core; thus what is referred to as `core mass' in this paper is
really the total heavy element excess over solar abundances. 
(6) The gas accretion rate is determined by the
requirement that the outer boundary of the protoplanet remain at the
modified accretion radius (eq.~2). Three-dimensional simulations
of a planet embedded in a disk \citep{lissauer09} determine in part
the region of a planet's gravitational influence.  (7) The opacity 
caused by grains in the envelope of the protoplanet is determined
by a detailed calculation of grain settling and coagulation, with   
the grain size distribution and the resulting grain opacity being 
recalculated at every level in the envelope at every time step.
It is this last assumption that represents a major improvement over
past calculations of Jupiter's formation, and it results in a
significant reduction in the formation time scale as compared to past
calculations. There are approximations involved in the treatment of
convection zones in the grain calculations, but the comparison of
cases $\sigma$10 and $\sigma$10R shows that these have a negligible effect
on the results.
The grains in the incoming gas are composed of silicates only, with an abundance by
mass of 1\% that of the gas.
Molecular opacities are included \citep{freedman08}.

The main parameter varied in this set of calculations is the
solid surface density $\sigma$ at 5.2 AU in the primordial disk; values
considered are 10, 6, and 4 g cm$^{-2}$. As in
previous simulations of Jupiter's formation using the prescription 
of \citet{lissauer87} for the runaway accretion of solids, 
the results show that there are
three main phases. Phase 1 involves primarily the accretion of solid
particles to build up the core; the mass of the gaseous envelope is
fairly negligible. Phase 2 begins when the isolation mass is reached; 
accretion of gas takes place at about three times the rate of
accretion of solid particles, and the envelope builds up in mass
relative to the core. Phase 3, rapid gas accretion, begins close to
the point where core and envelope masses are equal. The
calculations in this paper are taken beyond the  crossover mass
up to the point where the gas accretion rate becomes limited by
the rate at which the disk can supply the mass, as determined by
three-dimensional simulations ($1.5-3 \times 10^{-2}$ M$_\oplus$
yr$^{-1}$). The accretion up to the final mass of Jupiter takes place
quickly after that point \citep{lissauer09}.

The formation time to crossover is dominated by Phase 2 in all simulations.
Case $\sigma$10 may be compared with a very similar calculation
\citep{lissauer09} in which the grain opacity was simply
assumed to be 2\% that of interstellar grains. Phase 1 in the present case was completed in about the same
time as in the comparison case, but Phase 2 (0.52 Myr) was a factor of 3.6 shorter, 
giving a total formation time of about 1 Myr, compared with 2.6 Myr
in the comparison case.  Correspondingly,
the mean luminosity in phase 2 was a factor of 3.6 higher in the
present run than in the comparison case. The main cause of the increase
in luminosity is the very low opacity in the region around 600 K in
the present model (Fig.~\ref{fig:f3}), with a minimum of opacity of about $6 \times 10^{-4}$
cm$^2$ g$^{-1}$ as compared with $0.04$ cm$^2$ g$^{-1}$ at the same temperature in
the comparison case. Some layers in the present case have opacities
higher than 2\% interstellar, but they are mainly in the outermost layers
that are either convective or rather optically thin, so they do little
to influence the flow of radiation.
In spite of the difference in formation time for the two cases,
the core mass at onset of limiting gas accretion was practically 
the same in both cases, about 17 M$_\oplus$.  This mass is higher
than that usually derived for the present Jupiter \citep{saumon04}
but it is close to that predicted from an independent calculation of the
equation of state by \citet{militzer08}.

The question then arises whether a reduction in $\sigma$, which is
known to result in a smaller core mass, still leads to a reasonable
formation time. Cases $\sigma$6 and $\sigma$4 show that this is indeed
the case. The formation times are 1.9 and 4.0 Myr, respectively.
The ratios of Phase 2 times for the three cases scale as $M_{\rm iso}^{5/3}$/$L$, as
derived by \citet{pollack96} under the assumption that planetesimal
accretion provides most of the luminosity during Phase 2.
The value 4 Myr is  within the
 range of expected lifetimes of observed
circumstellar disks around young stars \citep{haisch01}.
Note that the $\sigma6$ case has a formation time only
14\% as long
(and Phase 2 time only 8\% as long) as the corresponding case in \citet{hubickyj05}
with grain opacities set at 2\%  of interstellar values, while the
$\sigma10$ case took 40\% as long as the corresponding case
($10\mathrm{L}\infty$) in that earlier work.

The main conclusions of this work are: (1) Within the basic framework
of a core-nucleated accretion model for Jupiter's formation,
 the effects of grain settling
and coagulation in the gaseous envelope of a protoplanet are 
significant in determining the gas accretion rate and formation time.
(2) The formation time for a Jupiter at 5.2 AU with a solid surface
density of 10 g cm$^{-2}$ (about 3 times that in the minimum mass
solar nebula) is about 1 Myr, significantly shorter than that in
comparable comparison cases. If $\sigma$ is taken to be 4 g cm$^{-2}$,
the formation time is
still reasonably short, about 4 Myr. (3) In the case with
$\sigma =  4$ g cm$^{-2}$, Jupiter's core mass is calculated to be
4.74 M$_\oplus$, consistent with the Jupiter structure models
computed by \citet{saumon04}. No arbitrary cutoff in the
core accretion rate was needed to obtain this result.

Further work is needed, however, to confirm this result. The main
points which need to be improved include, first, the planetesimal
accretion model, which should be replaced by a statistical
calculation taking into account the evolution of the  planetesimal size
distribution 
and the effects of the protoplanet itself on the planetesimal
velocity distribution \citep{weidenschilling97}. Numerous estimates
of the Phase 1 accretion time have been published. One example is
the (statistical + N-body) calculation of \citet{inaba03} which
takes into account planetesimal fragmentation as  well as the effect of
the gaseous envelope in increasing the effective cross-section. A 
planetesimal size of 10 km is assumed. If the opacity in the planetary
envelope is reduced by a factor 100 compared with interstellar
values,   a core approaching 10 M$_\oplus$
is reached in 5 Myr at 5.2 AU; however $\sigma$ is required to be about 5 times
that in the minimum mass solar nebula. This and other simulations, along
with the present results, suggest  that the formation time for Jupiter may be  
controlled by Phase 1, not Phase 2.
Second, the
effects of the ices should be included in the calculations of
grain  settling and coagulation. Third, the effects of the
dissolved accreted solid material, primarily ices, should be
included in the equation of state and opacity of the envelope
and in the calculation of the solid core mass.

\section*{Acknowledgments}
This work was supported in part by NASA Origins grant NNX08AH82G.
M.~P.~acknowledges the support of the Israel Academy of
Sciences. N.~M.~acknowledges
assistance from the Sackler Institute of Astronomy.

\bibliographystyle{elsarticle-harv}
\bibliography{astro-ph-10}

\begin{thebibliography}{33}
\expandafter\ifx\csname natexlab\endcsname\relax\def\natexlab#1{#1}\fi
\expandafter\ifx\csname url\endcsname\relax
  \def\url#1{\texttt{#1}}\fi
\expandafter\ifx\csname urlprefix\endcsname\relax\def\urlprefix{URL }\fi

\bibitem[{Alexander and Ferguson(1994)}]{alexander94}
Alexander, D.~R., Ferguson, J.~W., 1994. Low-temperature rosseland opacities.
  Astrophys.~J. 437, 879--891.

\bibitem[{Alibert et~al.(2005)Alibert, Mordasini, Benz, and
  Winisdoerffer}]{alibert05}
Alibert, Y., Mordasini, C., Benz, W., Winisdoerffer, C., 2005. Models of giant
  planet formation with migration and disk evolution. Astron.~Astrophys. 434,
  343--353.

\bibitem[{Bodenheimer et~al.(2000)Bodenheimer, Hubickyj, and
  Lissauer}]{bodenheimer00}
Bodenheimer, P., Hubickyj, O., Lissauer, J.~J., 2000. Models of the in situ
  formation of detected extrasolar giant planets. Icarus 143, 2--14.

\bibitem[{Bodenheimer and Pollack(1986)}]{bodenheimer86}
Bodenheimer, P., Pollack, J.~B., 1986. Calculations of the accretion and
  evolution of giant planets: the effects of solid cores. Icarus 67, 391--408.

\bibitem[{Cameron(1973)}]{cameron73}
Cameron, A.~G.~W., 1973. Accumulation processes in the primitive solar nebula.
  Icarus 18, 407--450.

\bibitem[{D'Angelo et~al.(2003)D'Angelo, Kley, and Henning}]{dangelo03}
D'Angelo, G., Kley, W., Henning, T., 2003. Orbital migration and mass accretion
  of protoplanets in three-dimensional global computations with nested grids.
  Astrophys.~J. 586, 540--561.

\bibitem[{Fortier et~al.(2007)Fortier, Bienvenuto, and Brunini}]{fortier07}
Fortier, A., Bienvenuto, O.~G., Brunini, A., 2007. Oligarchic planetesimal
  accretion and giant planet formation. Astron.~Astrophys. 473, 311--322.

\bibitem[{Freedman et~al.(2008)Freedman, Marley, and Lodders}]{freedman08}
Freedman, R.~S., Marley, M.~S., Lodders, K., 2008. Line and mean opacities for
  ultracool dwarfs and extrasolar planets. Astrophys.~J.~Suppl. 174, 504--513.

\bibitem[{Greenzweig and Lissauer(1992)}]{greenzweig92}
Greenzweig, Y., Lissauer, J.~J., 1992. Accretion rates of protoplanets. ii.
  gaussian distribution of planetesimal velocities. Icarus 100, 440--463.

\bibitem[{Haisch et~al.(2001)Haisch, Lada, and Lada}]{haisch01}
Haisch, Jr., K.~E., Lada, E.~A., Lada, C.~J., 2001. Disk frequencies and
  lifetimes in young clusters. Astrophys. J. 553, L153--L156.

\bibitem[{Henyey et~al.(1964)Henyey, Forbes, and Gould}]{henyey64}
Henyey, L., Forbes, J., Gould, N., 1964. A new method of automatic computation
  of stellar evolution. Astrophys.~J. 139, 306--317.

\bibitem[{Hubickyj et~al.(2005)Hubickyj, Bodenheimer, and
  Lissauer}]{hubickyj05}
Hubickyj, O., Bodenheimer, P., Lissauer, J.~J., 2005. Accretion of the gaseous
  envelope of jupiter around a 5--10 earth-mass core. Icarus 179, 415--431.

\bibitem[{Iaroslavitz and Podolak(2007)}]{Iaroslavitz07}
Iaroslavitz, E., Podolak, M., 2007. Atmospheric mass deposition by captured
  planetesimals. Icarus 187, 600--610.

\bibitem[{Ikoma et~al.(2000)Ikoma, Nakazawa, and Emori}]{ikoma00}
Ikoma, M., Nakazawa, K., Emori, H., 2000. Formation of giant planets:
  dependences on core accretion rate and grain opacity. Astrophys. J. 537,
  1013--1025.

\bibitem[{Inaba et~al.(2003)Inaba, Wetherill, and Ikoma}]{inaba03}
Inaba, S., Wetherill, G.~W., Ikoma, M., 2003. Formation of gas giant planets:
  core accretion models with fragmentation and planetary envelope. ICarus 166,
  46--62.

\bibitem[{Kokubo and Ida(2000)}]{kokubo00}
Kokubo, E., Ida, S., 2000. Formation of protoplanets from planetesimals in the
  solar nebula. Icarus 143, 15--27.

\bibitem[{Lissauer(1987)}]{lissauer87}
Lissauer, J.~J., 1987. Timescales for planetary accretion and the structure of
  the protoplanetary disk. ICarus 69, 249--265.

\bibitem[{Lissauer et~al.(2009)Lissauer, Hubickyj, D'Angelo, and
  Bodenheimer}]{lissauer09}
Lissauer, J.~J., Hubickyj, O., D'Angelo, G., Bodenheimer, P., 2009. Models of
  jupiter's growth incorporating thermal and hydrodynamic constraints. Icarus
  199, 338--350.

\bibitem[{Militzer et~al.(2008)Militzer, Hubbard, Vorberger, Tamblyn, and
  Bonev}]{militzer08}
Militzer, B., Hubbard, W.~B., Vorberger, J., Tamblyn, L., Bonev, S.~A., 2008. A
  massive core in jupiter predicted from first-principles simulations.
  Astrophys.~J. 688, L45--L48.

\bibitem[{Mizuno(1980)}]{mizuno80}
Mizuno, H., 1980. Formation of the giant planets. Prog.~Theor.~Phys. 64,
  544--557.

\bibitem[{Movshovitz and Podolak(2008)}]{movshovitz08}
Movshovitz, N., Podolak, M., 2008. The opacity of grains in protoplanetary
  atmospheres. Icarus 194, 368--378.

\bibitem[{Ormel and Cuzzi(2007)}]{ormel07}
Ormel, C.~W., Cuzzi, J.~N., 2007. Closed-form expressions for particle relative
  velocities induced by turbulence. Astron.~and Astrophys. 466, 413--420.

\bibitem[{Perri and Cameron(1974)}]{perri74}
Perri, F., Cameron, A.~G.~W., 1974. Hydrodynamic instability of the solar
  nebula in the presence of a planetary core. Icarus 22, 416--425.

\bibitem[{Podolak(2003)}]{podolak03}
Podolak, M., 2003. The contribution of small grains to the opacity of
  protoplanetary atmospheres. Icarus 165, 428--437.

\bibitem[{Podolak et~al.(1988)Podolak, Pollack, and Reynolds}]{podolak88}
Podolak, M., Pollack, J.~B., Reynolds, R.~T., 1988. Interactions of
  planetesimals with protoplanetary atmospheres. Icarus 73, 163--179.

\bibitem[{Pollack et~al.(1996)Pollack, Hubickyj, Bodenheimer, Lissauer,
  Podolak, and Greenzweig}]{pollack96}
Pollack, J.~B., Hubickyj, O., Bodenheimer, P., Lissauer, J.~J., Podolak, M.,
  Greenzweig, Y., 1996. Formation of the giant planets by concurrent accretion
  of solids and gas. Icarus 124, 62--85.

\bibitem[{Pollack et~al.(1985)Pollack, McKay, and Christofferson}]{pollack85}
Pollack, J.~B., McKay, C.~P., Christofferson, B., 1985. A calculation of the
  rosseland mean opacity of dust grains in primordial solar system nebulae.
  Icarus 64, 471--492.

\bibitem[{Safronov(1972)}]{safronov72}
Safronov, V.~S., 1972. Evolution of the protoplanetary cloud and formation of
  the earth and planets. In: Safronov, V.~S. (Ed.), Evolution of the
  protoplanetary cloud and formation of the earth and planets., by Safronov,
  V.~S..~ Translated from Russian.~Jerusalem (Israel): Israel Program for
  Scientific Translations. Keter Publishing House.

\bibitem[{Saumon et~al.(1995)Saumon, Chabrier, and van Horn}]{saumon95}
Saumon, D., Chabrier, G., van Horn, H., 1995. An equation of state for low-mass
  stars and giant planets. Astrophys.~J.~Suppl. 99, 713--741.

\bibitem[{Saumon and Guillot(2004)}]{saumon04}
Saumon, D., Guillot, T., 2004. Shock compression of deuterium and the interiors
  of jupiter and saturn. Astrophys. J. 609, 1170--1180.

\bibitem[{Stevenson(1982)}]{stevenson82}
Stevenson, D.~J., 1982. Formation of the giant planets. Planet.~Space Sci. 30,
  755--764.

\bibitem[{Weidenschilling(1984)}]{weidenschilling84}
Weidenschilling, S.~J., 1984. Evolution of grains in a turbulent solar nebula.
  Icarus 60, 553--567.

\bibitem[{Weidenschilling et~al.(1997)Weidenschilling, Spaute, Davis, Marzari,
  and Ohtsuki}]{weidenschilling97}
Weidenschilling, S.~J., Spaute, D., Davis, D.~R., Marzari, F., Ohtsuki, K.,
  1997. Accretional evolution of a planetesimal swarm. 2. the terrestrial zone.
  Icarus 128, 429--455.

\end{thebibliography}

\end{document}